\documentclass[reprint]{revtex4-1}
\usepackage{graphicx}
\usepackage{bm}
\usepackage{amsfonts}
\usepackage{mathtools}
\usepackage{amsmath}

\begin{document}


\title[Fourier analysis of the IR response of van der Waals materials]{Fourier analysis of the IR response of van der Waals materials}





\author{Anjan A. Reijnders}
\email[]{reijnder@physics.utoronto.ca}
\author{L.J. Sandilands}
\author{G. Pohl}
\author{K. W. Plumb}
\author{Young-June Kim}
\affiliation{Department of Physics \& Institute for Optical Sciences, University of Toronto, 60 St. George Street, Toronto, ON M5S 1A7, Canada}

\author{S. Jia}
\author{M.E. Charles}
\author{R.J. Cava}
\affiliation{Department of Chemistry, Princeton University, Princeton, NJ 08544, USA}

\author{K. S. Burch}
\email[]{ks.burch@bc.edu}
\affiliation{Department of Physics, Boston College, 140 Commonwealth Ave, Chestnut Hill, MA 02467, USA}

\date{\today}

\begin{abstract}
In this letter, we report on an analytical technique for optical investigations of semitransparent samples.
By Fourier transforming optical spectra with Fabry-P\'{e}rot resonances we extract information about sample thickness and its discrete variations. Moreover, this information is used to recover optical spectra devoid of Fabry-P\'{e}rot fringes, which simplifies optical modelling, and can reveal previously concealed spectral features. To illustrate its use, we apply our technique to a Si wafer as well as six different cleavable layered materials, including topological insulators, thermoelectrics, and magnetic insulators. In the layered materials, we find strong evidence of large step edges and thickness inhomogeneity, and cannot conclusively exclude the presence of voids in the bulk of cleaved samples. This could strongly affect the interpretation of transport and optical data of crystals with topologically protected surfaces states. 
\end{abstract}


\maketitle 
Since Geim and Novoselov's seminal paper\cite{Novoselov:2004it} on mechanically exfoliated graphite, a surge of renewed interest in cleavable layered materials hit the condensed matter physics and device engineering communities. As a cost effective alternative to molecular beam epitaxy, and without lattice matching constraints, mechanically exfoliated materials can be deposited on many substrates. Furthermore, many of these materials can be prepared in wafer scale sizes, suitable for industrial applications via chemical vapour deposition. \cite{Xu:2013hq,Geim:2013hf,Reina:2009cv,Park:2011em,vanderZande:2013cw,Butler:2013ha,Wang:2012ve,Kim:2008ed,Bonaccorso:2013iy} Using tape or razorblades to cleave these layered samples, clean and lustrous surfaces can be easily prepared for transport and optics experiments.\cite{Zhao:2011bs,Sandilands:2010tq,LaForge:2010dx,Tran:2014if,Xiong:2011wn,Forro:1990vt,Frindt:1965to,1972PhRvL..28..299F} While many groups have used this cleaving technique to prepare samples for experiments, little attention has been devoted to the integrity of the bulk crystals after cleaving. Particularly, a correct measurement of the sample thickness and surface quality can play a critical role in the analysis of both transport and optical data. 

Towards this end, we present a quick and easy optical technique to evaluate the sample thickness and surface quality of both carefully grown wafers as well mechanically cleaved layered samples that show Fabry-P\'{e}rot interference in their optical spectra. Moreover, this technique can be used to accurately remove Fabry-P\'{e}rot fringes from optical spectra. A common challenge in the analysis of optical spectroscopic data is the presence of Fabry-P\'{e}rot interference, obscuring  spectral features intrinsic to the sample's electronic response. Traditional approaches to circumvent this problem include data collection at reduced resolution (below the periodicity of Fabry-P\'{e}rot undulations), subtraction of Fabry-P\'{e}rot fringes by fitting optical spectra with sines and cosines, or simply regarding the Fabry-P\'{e}rot fringes as part of the sample's electronic response, resulting in optical constants with strong Fabry-P\'{e}rot fringes. In all scenarios it is clear that some information about the sample's intrinsic response is lost, and sharp spectral features cannot be resolved. A particularly relevant example is the analysis of thin films or exfoliated flakes on a thick substrate. The already weak optical response of the sample is often completely dominated by Fabry-P\'{e}rot  interference from the substrate. Currently, no good method exists for isolating the film or flake's response, which greatly limits the optical explorations of novel films and exfoliated flakes.

In this letter we show how this problem can be alleviated by exploiting Fabry-P\'{e}rot interference through Fourier analysis of optical spectra. The Fourier transform of broadband reflectance or transmittance data offers insights into the discrete thickness distribution of cleaved samples. This information is then used to remove Fabry-P\'{e}rot fringes from the optical spectra through an inverse Fourier transform of the filtered Fourier spectrum. We illustrate the efficacy of our approach by applying our technique to nine samples, including topological insulators, thermoelectrics, magnetic insulators, and semiconductors. Here we show how previously obscured spectral features can be resolved, and find that most of the cleaved materials show evidence of thickness inhomogeneity, which has important implications for the interpretations of experimental results.

%
\begin{figure}
\includegraphics[scale=0.34]{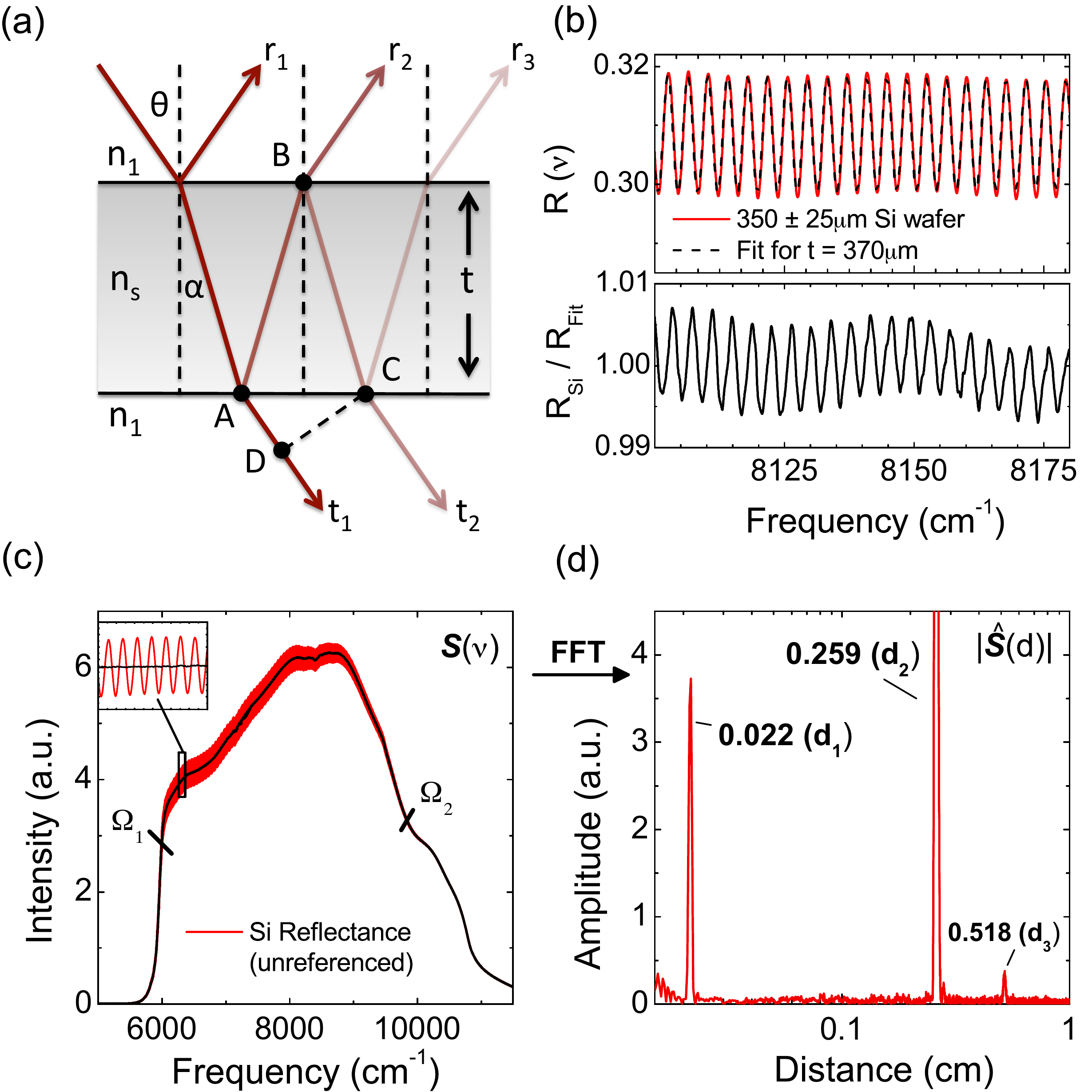}
\caption[Thin film interference geometry, Fabry-P\'{e}rot in optical spectra of a Si wafer, and a Fourier transform of the Reflectance intensity spectrum of a Si wafer]{\label{fig:FFTFig1}(a) Standard thin film geometry, showing how Fabry-P\'{e}rot occurs in both reflectance and transmittance spectra. (b) Top panel: Reflectance spectrum of a 350 $\pm$ 25 $\mu$m thick Si wafer (red), fitted by an optical model (black dashed line) revealing a thickness of 370 $\mu$m. Bottom panel: Despite the good fit, a division of R$_{Si}$ by R$_{Fit}$ exposes an additional weaker long range undulation, with a periodicity of $\sim1/10^{th}$ that of the Fabry-P\'{e}rot due to Si. This secondary Fabry-P\'{e}rot signal is associated with a polypropylene window. (c) Unreferenced reflectance spectrum of Si wafer, showing Fabry-P\'{e}rot fringes over the frequency range $\Omega_1-\Omega_2$ used for the FFT. (d) Amplitude of the FFT spectrum with distinct peaks associated with a 75 $\mu$m thick polypropylene window ($d_1$), and the $350\pm25 \mu$m thick Si wafer ($d_2$ and second harmonic $d_3$)}
\end{figure}
%

\section{Fabry-P\'{e}rot in optical spectra}
\label{sec:FP}
When a semi-transparent sample is placed in beam of coherent monochromatic electromagnetic radiation, and absorption is small, Fabry-P\'{e}rot interference can be observed in both the reflected and transmitted radiation. The interference condition can be easily deduced using Snell's law and geometry\cite{hecht}  (See Fig.~\ref{fig:FFTFig1}a) to be $\Lambda=2t\sqrt{n_s^2-n_1^2\sin^2\theta}$. Here, $\Lambda$ is the optical path length difference between rays in Fig.~\ref{fig:FFTFig1}a, $t$ is the sample thickness, $\theta$ is the incident angle, and $n_s$ and $n_1$ are the wavelength dependent real refractive indices of the sample and the medium surrounding the sample, respectively. Hence, whenever $\Lambda$ equals an integer multiple of the incident wavelength, the transmitted beam constructively interferes and peaks in intensity, while constructive interference for reflected beams occurs when $\Lambda$ equals half an odd integer multiple.

When a polychromatic source impinges on a sample, the reflectance and transmittance spectra will show Fabry-P\'{e}rot interference as a function of wavelength. These undulations can then be used to deduce the sample's thickness as follows. Starting with two close wavelengths $\lambda_a$ and $\lambda_b$, such that the complex refractive index $\tilde{n}(\lambda_a)\approx \tilde{n}(\lambda_b)$, we can take the difference between each wavelength's interference conditions and solve for $t$, such that
%
%
\begin{equation}\label{eq:FPt}
t=\frac{{\tfrac{1}{2}}M_{\nu_{b}-\nu_{a}}}{(\nu_a-\nu_b)\sqrt{n_{s}^2-n_1^2\sin^{2}\theta}}
\end{equation}
%
%
Here we have expressed wavelengths $\lambda_a$ and $\lambda_b$ in wavenumbers ($\nu=1/\lambda$) and $M_{\nu_{b}-\nu_{a}}$ is the number of peaks (or troughs) between $\nu_a$ and $\nu_b$. To illustrate the use of this method, the top panel of Fig.~\ref{fig:FFTFig1}b shows the reflectance spectrum of a piece of double side polished Si wafer, specified by the manufacturer to be $350\pm25 \mu$m thick (and $\rho= 20-30$  $\Omega$ cm). With $M_{\nu_{b}-\nu_{a}}=20$, $\nu_a-\nu_b=76.7$ cm$^{-1}$, $\theta=2.3^o$ (See Ref. \citenum{Reijnders:2014hq} for details of the experimental setup) and $n_s=3.517$\cite{Palik}, we find $t=371$ $\mu$m. This is consistent with optical modelling in RefFit\cite{KuzmenkoAB:2005jh} (black dashed line in Fig.~\ref{fig:FFTFig1}b), using the well documented complex refractive index of Si\cite{Palik}, which also yields a value of $370$ $\mu$m.

\section{Fourier Transform Analysis} 
\label{sec:FFTFP}
It is clear that eq.~\ref{eq:FPt} is an effective tool for the thickness determination of crystalline wafers, or samples with minimal thickness inhomogeneity. However, when layered materials are cleaved prior to measurements, smooth surfaces are difficult to prepare. Consequent beating between frequencies originating from sample regions of various thickness rule out the manual application of eq.~\ref{eq:FPt}. Moreover, optical modelling of spectra that show beating is extremely challenging, while experimental parameters (e.g. sample alignment, surface roughness, imperfect data normalization/referencing, etc.) further complicate the analysis. For example, the bottom panel of Fig.~\ref{fig:FFTFig1}b shows the reflectance of Si divided by its reflectance model. Despite the ostensibly successful fit, an additional long range undulation is observed, with a periodicity of $\sim1/10^{th}$ that of the Fabry-P\'{e}rot due to Si. This undulation is caused by Fabry-P\'{e}rot inside a polypropylene window in our setup (as we show later), which did not divide out perfectly when the reflectance intensity of Si was referenced with gold reflectance.

As a solution to this problem we propose the application of Fourier analysis of optical spectra. A similar approach has previously been used in the context of integrated optics, where Fourier analysis of optical transmission data through a  waveguide with a microcavity was used to determine the cavity's mode numbers.\cite{Blom:1997fj} Starting with any wavenumber dependent spectrum $\mathcal{S}(\nu)$ (e.g. Reflectance, Transmittance, or unreferenced Intensity), apodized between wavenumbers $\Omega_1$ and $\Omega_2$ such that all visible Fabry-P\'{e}rot interference fringes are included, a Fast Fourier Transform\cite{1969ITEdu..12...27C} 
%
%
\begin{equation}\label{eq:FT}
\hat{\mathcal{S}}(d)=\sum_{n=0}^{N-1}\mathcal{S}(\nu)~\mathrm{e}^{-2\pi i d \nu/N}
\end{equation}
%
%
yields a position ($d$) dependent complex quantity $\hat{\mathcal{S}}(d)$. Here $N$ corresponds to the number of equally spaced data points between $\Omega_1$ and $\Omega_2$ (i.e. the resolution of spectrum $\mathcal{S}(\nu)$). 
We note that $M_{\nu_{b}-\nu_{a}}/(\nu_a-\nu_b)$ is the inverse of the average spacing between fringes of one frequency in $\mathcal{S}(\nu)$ and thus corresponds to the position of maxima in the amplitude of $\hat{\mathcal{S}}(d)$. Hence, the peak position $d_p$ in $|\hat{\mathcal{S}}(d)|$ corresponds to a sample thickness of 
%
%
\begin{equation}\label{eq:FFTt}
t=d_p/2\sqrt{n_{s}^2-n_1^2\sin^{2}\theta}\end{equation}
%
%

To illustrate this approach, we turn to Si once more. Fig.~\ref{fig:FFTFig1}c shows the Intensity spectrum $I(\nu)$ (i.e. unreferenced reflectance) of the same $350\pm25$ $\mu$m thick Si wafer described before. The amplitude of the Fourier transformed $I(\nu)$ is shown in Fig.~\ref{fig:FFTFig1}d, revealing 3 distinct peak positions. Starting with $d_2=0.259$ cm, and $n_s=3.517$\cite{Palik}, we find $t=368$ $\mu$m, in excellent agreement with our previous results. The third peak at $d_3=0.518$ cm $=2d_2$ is a harmonic of $d_2$ (i.e. interference from higher order internal reflections such as $r_3$ in Fig.~\ref{fig:FFTFig1} a). Finally, the $d_1=0.022$ cm peak originates from a 75 $\mu$m thick polypropylene window used in the experimental setup. With $n_s=1.50$,\cite{Polypropylene,Birch:1992uc} we find $t=73$ $\mu$m, in good agreement with the manufacturer's specifications. We note that Fabry-P\'{e}rot interference from windows can, in principle, be removed by a reference measurement. However, nonlinear detectors, optical misalignment, and various other effects can result in residual Fabry-P\'{e}rot undulations even in corrected spectra. Hence, to illustrate how our technique can also register and correct such effects, we analyzed the unreferenced reflectance spectrum of Si.

Besides resolving the sample thickness (or $n_s$ if $t$ is already known), a major utility of the Fourier analysis approach is that the peaks in $|\hat{\mathcal{S}}(d)|$, associated with the sample thickness, can be easily removed, even without knowing $n_s$ or $t$. A subsequent inverse Fourier transform then yields the original optical spectrum, devoid of Fabry-P\'{e}rot interference, as shown by the black line in Fig.~\ref{fig:FFTFig1}c. This can greatly enhance the quality of optical fits, and reveals spectral features previously obscured by strong undulations. Moreover, even systemic sources of Fabry-P\'{e}rot fringes, such as windows, polarizers, or other optical elements, can be easily identified and removed, as illustrated in the Si example. 

\section{Results and discussion}
\label{sec:FFTresults}
Before applying this technique to cleaved van der Waals bound materials, it is useful to discuss what can be expected. One of the underlying assumptions of the described approach is that absorption is minimal and $n_s$ is constant over the chosen spectral range. Since $\tilde{n}=\sqrt{\tilde{\epsilon}}$, where $\tilde{\epsilon}=\epsilon_1+i\epsilon_2$ is the complex dielectric function, we have $n_s=(\tfrac{\sqrt{\epsilon_1^2+\epsilon_2^2}+\epsilon_1}{2})^{1/2}$. Hence, as long as $\epsilon_1\gg\epsilon_2$ (i.e. minimal absorption) and $\mathrm{d}\epsilon_1/\mathrm{d}\nu\approx$ constant (i.e. $n_s$ is constant), the Fourier transform approach works very well. However, while this approximation generally gains validity as $(\Omega_2-\Omega_1)\rightarrow 0$, the smallest resolvable spectral feature of the Fourier transform is inversely proportional to $(\Omega_2-\Omega_1)$.\cite{davis2001fourier} A balance must thus be sought where a sufficiently large spectral range is chosen that includes all spacial frequencies associated with expected sample thicknesses. Hence, the inevitable finite dispersion of $\tilde{n}$ across the selected spectral range will cause the peaks in $|\hat{\mathcal{S}}(d)|$ to broaden. 
We note that peak broadening also occurs for samples with a smooth thickness gradient. Indeed, this is commonly exploited in optical measurements, where deliberately wedged samples are used to suppress Fabry-P\'{e}rot fringes. In mechanically cleaved samples, however, such a smooth gradient is unlikely.\\

%
\begin{figure}[]
\includegraphics[scale=0.79]{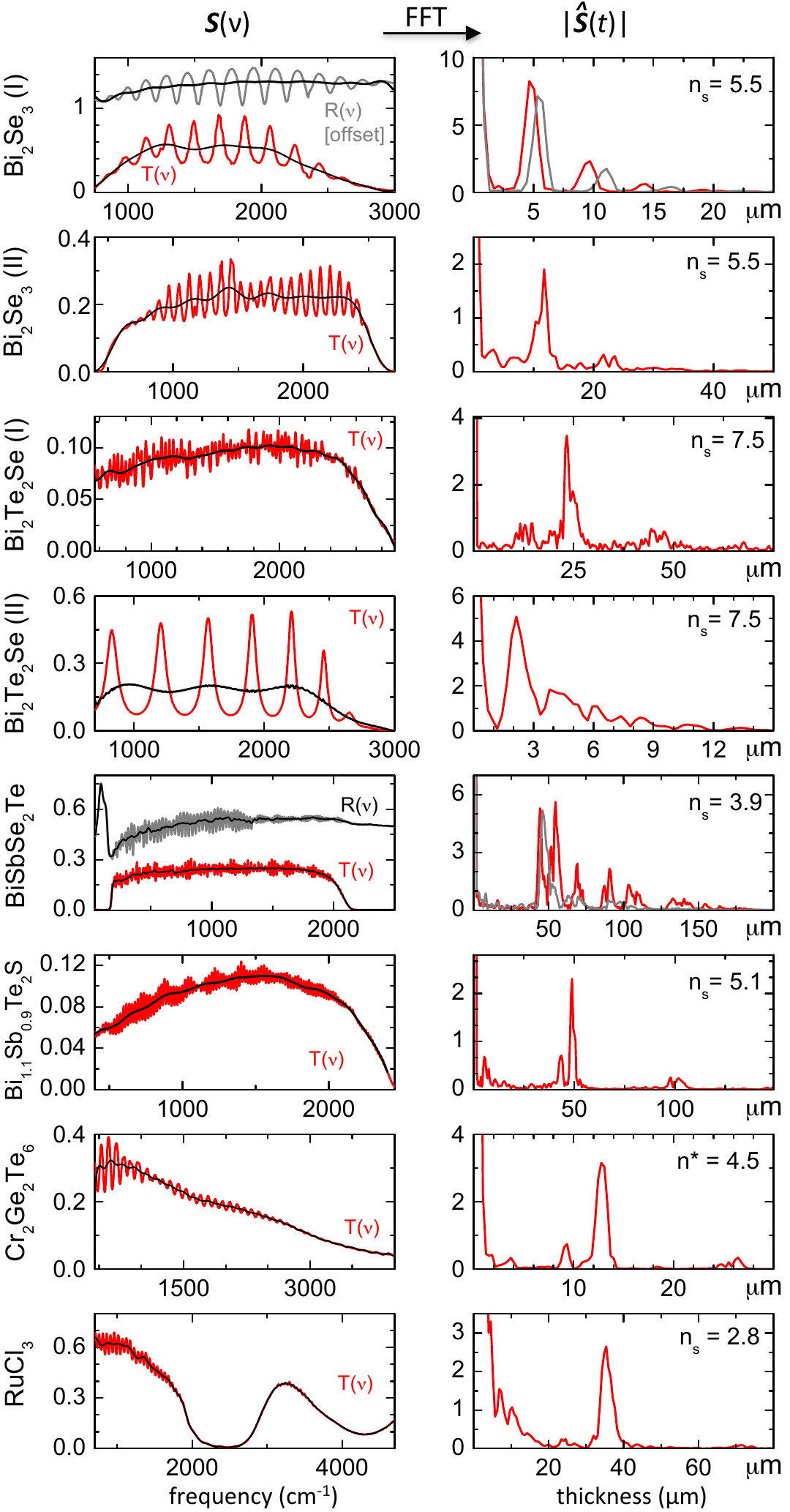}
\caption[Optical reflectance or transmittance spectra of Bi$_2$Se$_3$, (In$_{0.88}$Bi$_{0.12}$)$_{2}$Se$_{3}$, Bi$_2$Te$_2$Se, BiSbSe$_2$Te, Bi$_{1.1}$Sb$_{0.9}$Te$_2$S, Cr$_2$Ge$_2$Te$_6$, and RuCl$_3$ and their Fourier transform]{\label{fig:FFTFig2}
Optical spectra $\mathcal{S}(\nu)$ (left) and their Fourier transform $\hat{\mathcal{S}}(t)$ (right) for various weakly van der Waals bound compounds, revealing distinct peaks (and harmonics) at the sample thickness. The black line in the left panel is the inverse Fourier transform of $\hat{\mathcal{S}}(t)$ after removal of the peaks associated with the sample thickness.}
\end{figure}
%
Fig.~\ref{fig:FFTFig2} shows the optical spectra (left panel) of six different materials, where separate measurements on samples with identical stoichiometry are designated by a I or II.
With the exception of both Bi$_2$Se$_3$ samples and the Bi$_2$Te$_2$Se (II) sample, all spectra were recorded at the University of Toronto using a modified Bruker Vertex 80v FTIR interferometer, described elsewhere.\cite{Reijnders:2014hq} Spectra of Bi$_2$Se$_3$ and Bi$_2$Te$_2$Se (II) were digitized from previous publications.\cite{LaForge:2010dx,Segura:2012wb,2012arXiv1209.3593A} All Bi-containing samples and Cr$_2$Ge$_2$Te$_6$ were grown at Princeton University,\cite{Xiong:2011tm,Ji:2013dg} and RuCl$_3$ was grown in Toronto.\cite{Plumb:2014tv} Prior to reflectance or transmittance measurements, samples were mechanically cleaved using tape or were measured as-grown (RuCl$_3$), with typical thickness $<150$ $\mu$m. The right panel shows the Fourier transform of the optical spectra. Since it is easier to work with a thickness distribution $|\hat{\mathcal{S}}(t)|$  rather than an {\it optical} thickness distribution $|\hat{\mathcal{S}}(d)|$, we used eq.~\ref{eq:FFTt} to plot all spectra with thickness in microns on the $x$-axis. To do so, we assumed that $\theta=0^o$ and $n_1=1$ for the digitized Bi$_2$Se$_3$ and Bi$_2$Te$_2$Se (II) spectra, and estimated $n_s=4.5$ (the measured value at 8000 cm$^{-1}$) for Cr$_2$Ge$_2$Te$_6$. Real refractive index values for all other samples were measured in Toronto,\cite{Reijnders3,Reijnders:2014hq,Sandilands3} or reported in the literature.\cite{KOHLER:1974tf}

\begin{figure}[]
\includegraphics[scale=0.79]{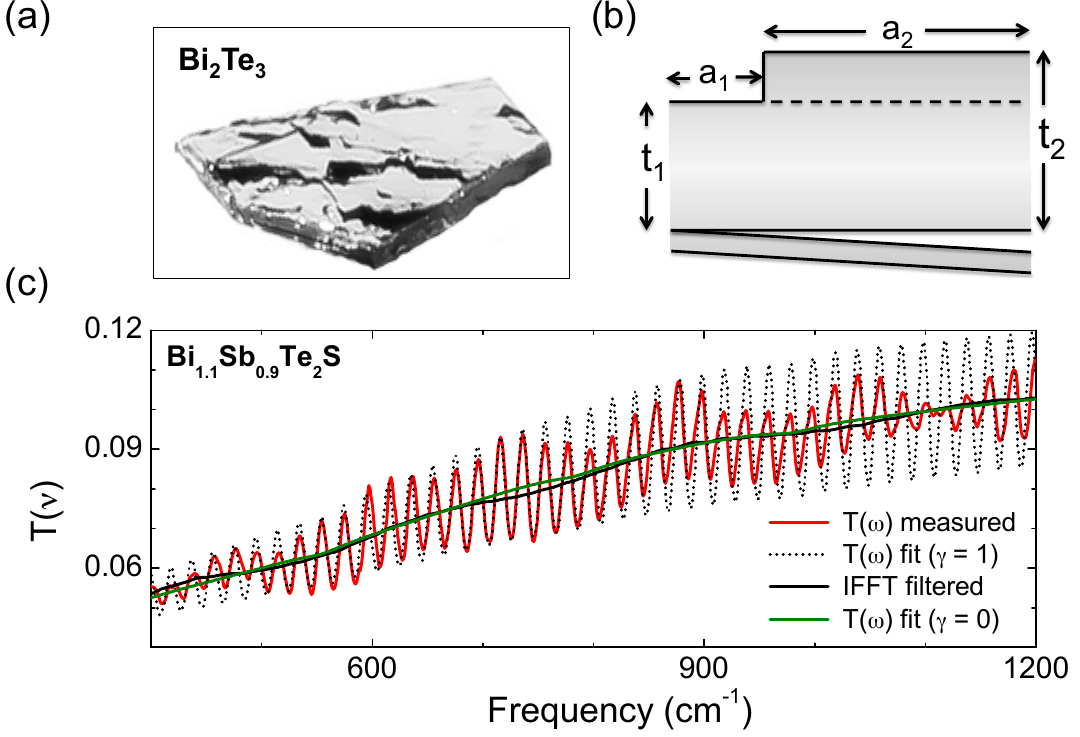}
\caption[Photograph and model of Topological insulator surface step edges and flakes, and transmittance spectrum of Bi$_{1.1}$Sb$_{0.9}$Te$_2$S showing the FFT method reproduces optical data without Fabry-P\'{e}rot fringes.]{\label{fig:FFTFig3}
(a) Surface of a poorly cleaved Bi$_2$Te$_3$ crystal. (b) Layered materials can show both step edges and/or partially detached flakes. Samples with step edges can be optically described as an effective medium with various thickness (e.g. $t_1$ and $t_2$) with associated volume fractions (i.e. functions of areas $a_1$ and $a_2$). (c) Transmittance spectrum and optical model of Bi$_{1.1}$Sb$_{0.9}$Te$_2$S. When optical coherence $\gamma$ in the fit is set to 0, the transmittance spectrum matches the filtered IFFT of the Bi$_{1.1}$Sb$_{0.9}$Te$_2$S thickness distribution $|\hat{\mathcal{S}}(t)|$. This confirms that only Fabry-P\'{e}rot fringes were removed, and all intrinsic spectral features are preserved.}
\end{figure}

Most notably, all $|\hat{\mathcal{S}}(t)|$ plots show a single dominant peak accompanied by various smaller peaks, where the dominant peak is associated with the most prominent sample thickness. To further confirm this interpretation, the thickness of both BiSbSe$_2$Te and Bi$_{1.1}$Sb$_{0.9}$Te$_2$S crystals were independently determined with a (mechanical) digital thickness gauge, and found to be $50\pm5$ $\mu$m for both samples (as predicted by our Fourier analysis shown in Fig.\ref{fig:FFTFig2}). The smaller peaks in $|\hat{\mathcal{S}}(t)|$ either result from discrete sample thickness variations (e.g. 14 $\mu$m for Cr$_2$Ge$_2$Te$_6$), or are due to harmonics as observed in Si (e.g. 100 $\mu$m = $2\times50$ $\mu$m for Bi$_{1.1}$Sb$_{0.9}$Te$_2$S, or 9 $\mu$m = $2\times4.5$ $\mu$m, and 13.5 $\mu$m = $3\times4.5$ $\mu$m for Bi$_2$Se$_3$ (I)). 
Such discrete sample thickness variations (i.e. inhomogeneity in the $xy$-plane) can be easily understood by inspecting the surface of (poorly) cleaved layered materials, such as Bi$_2$Te$_3$ shown in Fig.~\ref{fig:FFTFig3}a. These samples can be modelled as an effective medium, where the total measured optical response ${\mathcal{S}}(\nu)$ is a sum of volume fractions multiplied by their respective optical response function. For example, the measured reflectance $R_m$ of a sample with two regions of different thickness (as shown in the top half of Fig.~\ref{fig:FFTFig3} b) would be $R_m=f_1R_1+(1-f_1)R_2$. Here, $f_1$ and $(1-f_1)$ are a function of volume fractions $a_1t_1$ and $a_2t_2$, respectively, and $R_i$ are the squared magnitudes of the associated complex Fresnel reflection coefficients. 

We note that inhomogeneity in the $z$-direction, such as voids inside a crystal (as shown in the bottom half of Fig.~\ref{fig:FFTFig3} b) could also produce multiple peaks in $|\hat{\mathcal{S}}(t)|$. However, in such a scenario the total optical response would be given by the squared norm of the sum of Fresnel coefficients (e.g. $R=|\sum_{i=1}^{N}r_i|^2$ where $r_i$ are reflection coefficients for each interface in the multilayered sample). In other words, for inhomogeneity along $z$, the complex amplitudes are added, as opposed to intensities for $xy$ inhomogeneities.  Therefore, cross terms in $R$ result in sums and differences of the various Fabry-P\'{e}rot frequencies. Although we see no obvious signatures of this effect in Fig.~\ref{fig:FFTFig2}, it can be used as a discriminating technique in determining sample integrity. We note that detection of such additional thin layers may be difficult since the typical resolution of $|\hat{\mathcal{S}}(t)|$ is $\geq1\mu$m. Hence, it is possible that the observed width of the peaks in $|\hat{\mathcal{S}}(t)|$, even when $\mathrm{d}\epsilon_1/\mathrm{d}\nu\approx$ constant, is a result of unresolved peaks from inhomogeneities in the $z$ direction, symmetrically positioned below and above the dominant peak. Regardless, from the many peaks in $|\hat{\mathcal{S}}(t)|$ (and Fig.~\ref{fig:FFTFig3}a) it is clear that none of the layered materials produce an optically flat surface across the full sample when cleaved. The observation of Fabry-P\'{e}rot fringes, however, does imply high surface quality for all individual regions of constant thickness. 

In principle, all peaks can be fitted so that their center position, width and spectral weight can be used to replicate the measured Fabry-P\'{e}rot fringes in an optical model. More convenient, however, is to remove the appropriate peaks from $|\hat{\mathcal{S}}(t)|$, and fit its inverse Fourier transform spectrum with a model that ignores Fabry-P\'{e}rot. The black lines in the left panels of Fig.~\ref{fig:FFTFig2} show the result of such optical spectra, in which all Fabry-P\'{e}rot fringes have been removed. To confirm that such filtering only removes the Fabry-P\'{e}rot fringes, and preserves all other spectral information, the transmittance of Bi$_{1.1}$Sb$_{0.9}$Te$_2$S was fit with a model that captures the Fabry-P\'{e}rot fringes reasonably well, as shown in Fig.~\ref{fig:FFTFig3}b. With the degree of coherence ($\gamma$) being a variable in our model, the Fabry-P\'{e}rot fringes were artificially suppressed ($\gamma=0$), as shown by the green line in Fig.~\ref{fig:FFTFig3}b. It is clear that the FFT filtered spectrum (black line) matches the optical model with $\gamma=0$ very well, thus confirming that all spectral information but the Fabry-P\'{e}rot fringes are preserved.\\
%
%

A comparison of Bi$_{1.1}$Sb$_{0.9}$Te$_2$S ($t\approx50$ $\mu$m), Bi$_2$Te$_2$Se I ($t\approx24$ $\mu$m), Bi$_2$Se$_3$ II ($t\approx12$ $\mu$m), and Bi$_2$Te$_2$Se II ($t\approx2$ $\mu$m), shows that as samples get thinner, it becomes increasingly difficult to distinguish thickness related Fabry-P\'{e}rot from features due to intrinsic electronic structure. This is particularly clear for RuCl$_3$, where samples with $t<10$ $\mu$m would produce Fabry-P\'{e}rot fringes that are hard to distinguish from intrinsic electronic absorption. Nevertheless, a major advantage of this Fourier transform filter technique is that it addresses only specific frequencies associated with Fabry-P\'{e}rot interference, while no knowledge of the refractive index is required.

Finally, we discuss the comparison of reflectance and transmittance spectra on the same sample, as measured for Bi$_2$Se$_3$ (I) and BiSbSe$_2$Te. The interference condition described in section~\ref{sec:FP} predicts that for identical values for $\theta$ and $n_1$, the Fabry-P\'{e}rot fringes of reflectance and transmittance spectra should overlap, so that $|\hat{\mathcal{S}}(t)|_{R(\nu)}=|\hat{\mathcal{S}}(t)|_{T(\nu)}$. However, from Fig.~\ref{fig:FFTFig2} it is clear that this is not always the case. While we cannot comment on the experimental conditions under which the Bi$_2$Se$_3$ (I) spectra were obtained,\cite{Segura:2012wb} it is possible that the sample was cleaved between $R(\nu)$ and $T(\nu)$ measurements (thus reducing the thickness), or the incident angle $\theta$ changed between measurements. For BiSbSe$_2$Te, we note that the unique setup in Toronto\cite{Reijnders:2014hq} allows $\theta$ and $n_1$ to be identical for both $R(\nu)$ and $T(\nu)$ measurements on a single sample (i.e. thickness remains constant too). The presence of additional peaks in the $T(\nu)$ spectra, not visible in $R(\nu)$ could indicate the presence of (partially) detached flakes on the back surface of the sample, to which only transmission measurements are sensitive. Such a scenario is illustrated in Fig.~\ref{fig:FFTFig3}b, and occasionally observed upon careful visual inspection of the sample's surface when the flake is sufficiently detached from the bulk crystal.  While this can only be confirmed with alternative techniques (e.g. x-ray diffraction or Scanning Electron Microscopy), it would be of considerable relevance. In topological insulators, such as BiSbSe$_2$Te, multiple surfaces would result in a greater surface state contribution to optical/electronic properties; a scenario that has been previously suggested.\cite{2012arXiv1209.3593A,Jenkins:2010kg,Cao:2012kc}
An alternative explanation for the mismatch between $|\hat{\mathcal{S}}(t)|_{R(\nu)}$ and $|\hat{\mathcal{S}}(t)|_{T(\nu)}$, however, is the imperfect overlap of the probed sample regions between $R(\nu)$ and $T(\nu)$, as a result of misalignment, beam clipping, or chromatic aberrations.

\section{Conclusion}
\label{sec:FFTconclusion}
In this letter, we have shown how Fourier analysis of optical spectra offers some previously unexplored opportunities in the study of novel materials. When the real refractive index of a material is known, this technique provides a quick measure of the sample thickness and discrete thickness variations of semi-transparent samples, without the application of a physical (potentially destructive) probe. Moreover, even without knowledge of the refractive index, the inverse Fourier transform of a filtered thickness spectrum yields an optical spectrum devoid of Fabry-P\'{e}rot fringes. Such treatment can expose previously obscured features, and can also be used to remove the strong optical effects of sample substrates, windows, or other optical components that cause Fabry-P\'{e}rot interference.

We have shown how this technique is applicable not only to carefully manufactured semiconductors, such as Si wafers, but also to a wide variety of cleavable materials, including topological insulators, thermoelectrics, and magnetic insulators. Hence, with recent renewed interest in novel layered samples, we believe this technique makes a valuable addition to standard optical analysis techniques.

\begin{acknowledgments}
This work has been partially funded by the Ontario Research Fund, the Natural Sciences and Engineering Research council of Canada, and the Canada Foundation for Innovation. The crystal growth work at Princeton university was supported by the NSF MRSEC grant DMR 0819860 (topological insulators) and AFOSR MURI grant FA9550-10-1-0553 (thermoelectrics). \end{acknowledgments}


\begin{thebibliography}{37}%
\makeatletter
\providecommand \@ifxundefined [1]{%
 \@ifx{#1\undefined}
}%
\providecommand \@ifnum [1]{%
 \ifnum #1\expandafter \@firstoftwo
 \else \expandafter \@secondoftwo
 \fi
}%
\providecommand \@ifx [1]{%
 \ifx #1\expandafter \@firstoftwo
 \else \expandafter \@secondoftwo
 \fi
}%
\providecommand \natexlab [1]{#1}%
\providecommand \enquote  [1]{``#1''}%
\providecommand \bibnamefont  [1]{#1}%
\providecommand \bibfnamefont [1]{#1}%
\providecommand \citenamefont [1]{#1}%
\providecommand \href@noop [0]{\@secondoftwo}%
\providecommand \href [0]{\begingroup \@sanitize@url \@href}%
\providecommand \@href[1]{\@@startlink{#1}\@@href}%
\providecommand \@@href[1]{\endgroup#1\@@endlink}%
\providecommand \@sanitize@url [0]{\catcode `\\12\catcode `\$12\catcode
  `\&12\catcode `\#12\catcode `\^12\catcode `\_12\catcode `\%12\relax}%
\providecommand \@@startlink[1]{}%
\providecommand \@@endlink[0]{}%
\providecommand \url  [0]{\begingroup\@sanitize@url \@url }%
\providecommand \@url [1]{\endgroup\@href {#1}{\urlprefix }}%
\providecommand \urlprefix  [0]{URL }%
\providecommand \Eprint [0]{\href }%
\providecommand \doibase [0]{http://dx.doi.org/}%
\providecommand \selectlanguage [0]{\@gobble}%
\providecommand \bibinfo  [0]{\@secondoftwo}%
\providecommand \bibfield  [0]{\@secondoftwo}%
\providecommand \translation [1]{[#1]}%
\providecommand \BibitemOpen [0]{}%
\providecommand \bibitemStop [0]{}%
\providecommand \bibitemNoStop [0]{.\EOS\space}%
\providecommand \EOS [0]{\spacefactor3000\relax}%
\providecommand \BibitemShut  [1]{\csname bibitem#1\endcsname}%
\let\auto@bib@innerbib\@empty
\bibitem [{\citenamefont {Novoselov}\ \emph {et~al.}(2004)\citenamefont
  {Novoselov}, \citenamefont {Geim}, \citenamefont {Morozov}, \citenamefont
  {Jiang}, \citenamefont {Zhang}, \citenamefont {Dubonos}, \citenamefont
  {Grigorieva},\ and\ \citenamefont {Firsov}}]{Novoselov:2004it}%
  \BibitemOpen
  \bibfield  {author} {\bibinfo {author} {\bibfnamefont {K.~S.}\ \bibnamefont
  {Novoselov}}, \bibinfo {author} {\bibfnamefont {A.~K.}\ \bibnamefont {Geim}},
  \bibinfo {author} {\bibfnamefont {S.~V.}\ \bibnamefont {Morozov}}, \bibinfo
  {author} {\bibfnamefont {D.}~\bibnamefont {Jiang}}, \bibinfo {author}
  {\bibfnamefont {Y.}~\bibnamefont {Zhang}}, \bibinfo {author} {\bibfnamefont
  {S.~V.}\ \bibnamefont {Dubonos}}, \bibinfo {author} {\bibfnamefont {I.~V.}\
  \bibnamefont {Grigorieva}}, \ and\ \bibinfo {author} {\bibfnamefont {A.~A.}\
  \bibnamefont {Firsov}},\ }\href@noop {} {\bibfield  {journal} {\bibinfo
  {journal} {Science}\ }\textbf {\bibinfo {volume} {306}},\ \bibinfo {pages}
  {666} (\bibinfo {year} {2004})}\BibitemShut {NoStop}%
\bibitem [{\citenamefont {Xu}\ \emph {et~al.}(2013)\citenamefont {Xu},
  \citenamefont {Liang}, \citenamefont {Shi},\ and\ \citenamefont
  {Chen}}]{Xu:2013hq}%
  \BibitemOpen
  \bibfield  {author} {\bibinfo {author} {\bibfnamefont {M.}~\bibnamefont
  {Xu}}, \bibinfo {author} {\bibfnamefont {T.}~\bibnamefont {Liang}}, \bibinfo
  {author} {\bibfnamefont {M.}~\bibnamefont {Shi}}, \ and\ \bibinfo {author}
  {\bibfnamefont {H.}~\bibnamefont {Chen}},\ }\href@noop {} {\bibfield
  {journal} {\bibinfo  {journal} {Chemical Reviews}\ }\textbf {\bibinfo
  {volume} {113}},\ \bibinfo {pages} {3766} (\bibinfo {year}
  {2013})}\BibitemShut {NoStop}%
\bibitem [{\citenamefont {Geim}\ and\ \citenamefont
  {Grigorieva}(2013)}]{Geim:2013hf}%
  \BibitemOpen
  \bibfield  {author} {\bibinfo {author} {\bibfnamefont {A.~K.}\ \bibnamefont
  {Geim}}\ and\ \bibinfo {author} {\bibfnamefont {I.~V.}\ \bibnamefont
  {Grigorieva}},\ }\href@noop {} {\bibfield  {journal} {\bibinfo  {journal}
  {Nature}\ }\textbf {\bibinfo {volume} {499}},\ \bibinfo {pages} {419}
  (\bibinfo {year} {2013})}\BibitemShut {NoStop}%
\bibitem [{\citenamefont {Reina}\ \emph {et~al.}(2009)\citenamefont {Reina},
  \citenamefont {Jia}, \citenamefont {Ho}, \citenamefont {Nezich},
  \citenamefont {Son}, \citenamefont {Bulovic}, \citenamefont {Dresselhaus},\
  and\ \citenamefont {Kong}}]{Reina:2009cv}%
  \BibitemOpen
  \bibfield  {author} {\bibinfo {author} {\bibfnamefont {A.}~\bibnamefont
  {Reina}}, \bibinfo {author} {\bibfnamefont {X.}~\bibnamefont {Jia}}, \bibinfo
  {author} {\bibfnamefont {J.}~\bibnamefont {Ho}}, \bibinfo {author}
  {\bibfnamefont {D.}~\bibnamefont {Nezich}}, \bibinfo {author} {\bibfnamefont
  {H.}~\bibnamefont {Son}}, \bibinfo {author} {\bibfnamefont {V.}~\bibnamefont
  {Bulovic}}, \bibinfo {author} {\bibfnamefont {M.~S.}\ \bibnamefont
  {Dresselhaus}}, \ and\ \bibinfo {author} {\bibfnamefont {J.}~\bibnamefont
  {Kong}},\ }\href@noop {} {\bibfield  {journal} {\bibinfo  {journal} {Nano
  Letters}\ }\textbf {\bibinfo {volume} {9}},\ \bibinfo {pages} {30} (\bibinfo
  {year} {2009})}\BibitemShut {NoStop}%
\bibitem [{\citenamefont {Park}(2011)}]{Park:2011em}%
  \BibitemOpen
  \bibfield  {author} {\bibinfo {author} {\bibfnamefont {J.-U.}\ \bibnamefont
  {Park}},\ }\href@noop {} {\bibfield  {journal} {\bibinfo  {journal} {Nature
  Materials}\ }\textbf {\bibinfo {volume} {11}},\ \bibinfo {pages} {120}
  (\bibinfo {year} {2011})}\BibitemShut {NoStop}%
\bibitem [{\citenamefont {van~der Zande}(2013)}]{vanderZande:2013cw}%
  \BibitemOpen
  \bibfield  {author} {\bibinfo {author} {\bibfnamefont {A.~M.}\ \bibnamefont
  {van~der Zande}},\ }\href@noop {} {\bibfield  {journal} {\bibinfo  {journal}
  {Nature Materials}\ }\textbf {\bibinfo {volume} {12}},\ \bibinfo {pages}
  {554} (\bibinfo {year} {2013})}\BibitemShut {NoStop}%
\bibitem [{\citenamefont {Butler}\ \emph {et~al.}(2013)\citenamefont {Butler},
  \citenamefont {Hollen}, \citenamefont {Cao}, \citenamefont {Cui},
  \citenamefont {Gupta}, \citenamefont {Guti{\'e}rrez}, \citenamefont {Heinz},
  \citenamefont {Hong}, \citenamefont {Huang}, \citenamefont {Ismach},
  \citenamefont {Johnston-Halperin}, \citenamefont {Kuno}, \citenamefont
  {Plashnitsa}, \citenamefont {Robinson}, \citenamefont {Ruoff}, \citenamefont
  {Salahuddin}, \citenamefont {Shan}, \citenamefont {Shi}, \citenamefont
  {Spencer}, \citenamefont {Terrones}, \citenamefont {Windl},\ and\
  \citenamefont {Goldberger}}]{Butler:2013ha}%
  \BibitemOpen
  \bibfield  {author} {\bibinfo {author} {\bibfnamefont {S.~Z.}\ \bibnamefont
  {Butler}}, \bibinfo {author} {\bibfnamefont {S.~M.}\ \bibnamefont {Hollen}},
  \bibinfo {author} {\bibfnamefont {L.}~\bibnamefont {Cao}}, \bibinfo {author}
  {\bibfnamefont {Y.}~\bibnamefont {Cui}}, \bibinfo {author} {\bibfnamefont
  {J.~A.}\ \bibnamefont {Gupta}}, \bibinfo {author} {\bibfnamefont {H.~R.}\
  \bibnamefont {Guti{\'e}rrez}}, \bibinfo {author} {\bibfnamefont {T.~F.}\
  \bibnamefont {Heinz}}, \bibinfo {author} {\bibfnamefont {S.~S.}\ \bibnamefont
  {Hong}}, \bibinfo {author} {\bibfnamefont {J.}~\bibnamefont {Huang}},
  \bibinfo {author} {\bibfnamefont {A.~F.}\ \bibnamefont {Ismach}}, \bibinfo
  {author} {\bibfnamefont {E.}~\bibnamefont {Johnston-Halperin}}, \bibinfo
  {author} {\bibfnamefont {M.}~\bibnamefont {Kuno}}, \bibinfo {author}
  {\bibfnamefont {V.~V.}\ \bibnamefont {Plashnitsa}}, \bibinfo {author}
  {\bibfnamefont {R.~D.}\ \bibnamefont {Robinson}}, \bibinfo {author}
  {\bibfnamefont {R.~S.}\ \bibnamefont {Ruoff}}, \bibinfo {author}
  {\bibfnamefont {S.}~\bibnamefont {Salahuddin}}, \bibinfo {author}
  {\bibfnamefont {J.}~\bibnamefont {Shan}}, \bibinfo {author} {\bibfnamefont
  {L.}~\bibnamefont {Shi}}, \bibinfo {author} {\bibfnamefont {M.~G.}\
  \bibnamefont {Spencer}}, \bibinfo {author} {\bibfnamefont {M.}~\bibnamefont
  {Terrones}}, \bibinfo {author} {\bibfnamefont {W.}~\bibnamefont {Windl}}, \
  and\ \bibinfo {author} {\bibfnamefont {J.~E.}\ \bibnamefont {Goldberger}},\
  }\href@noop {} {\bibfield  {journal} {\bibinfo  {journal} {ACS Nano}\
  }\textbf {\bibinfo {volume} {7}},\ \bibinfo {pages} {2898} (\bibinfo {year}
  {2013})}\BibitemShut {NoStop}%
\bibitem [{\citenamefont {Wang}\ \emph {et~al.}(2012)\citenamefont {Wang},
  \citenamefont {Yu}, \citenamefont {Lee}, \citenamefont {Fang}, \citenamefont
  {Hsu}, \citenamefont {Herring}, \citenamefont {Chin}, \citenamefont {Dubey},
  \citenamefont {Li},\ and\ \citenamefont {Kong}}]{Wang:2012ve}%
  \BibitemOpen
  \bibfield  {author} {\bibinfo {author} {\bibfnamefont {H.}~\bibnamefont
  {Wang}}, \bibinfo {author} {\bibfnamefont {L.}~\bibnamefont {Yu}}, \bibinfo
  {author} {\bibfnamefont {Y.-H.}\ \bibnamefont {Lee}}, \bibinfo {author}
  {\bibfnamefont {W.}~\bibnamefont {Fang}}, \bibinfo {author} {\bibfnamefont
  {A.}~\bibnamefont {Hsu}}, \bibinfo {author} {\bibfnamefont {P.}~\bibnamefont
  {Herring}}, \bibinfo {author} {\bibfnamefont {M.}~\bibnamefont {Chin}},
  \bibinfo {author} {\bibfnamefont {M.}~\bibnamefont {Dubey}}, \bibinfo
  {author} {\bibfnamefont {L.-J.}\ \bibnamefont {Li}}, \ and\ \bibinfo {author}
  {\bibfnamefont {J.}~\bibnamefont {Kong}},\ }\href@noop {} {\bibfield
  {journal} {\bibinfo  {journal} {IEEE}\ ,\ \bibinfo {pages} {IEDM12}}
  (\bibinfo {year} {2012})}\BibitemShut {NoStop}%
\bibitem [{\citenamefont {Kim}\ \emph {et~al.}(2008)\citenamefont {Kim},
  \citenamefont {Zhao}, \citenamefont {Jang}, \citenamefont {Lee},
  \citenamefont {Kim}, \citenamefont {Kim}, \citenamefont {Ahn}, \citenamefont
  {Kim}, \citenamefont {Choi},\ and\ \citenamefont {Hong}}]{Kim:2008ed}%
  \BibitemOpen
  \bibfield  {author} {\bibinfo {author} {\bibfnamefont {K.~S.}\ \bibnamefont
  {Kim}}, \bibinfo {author} {\bibfnamefont {Y.}~\bibnamefont {Zhao}}, \bibinfo
  {author} {\bibfnamefont {H.}~\bibnamefont {Jang}}, \bibinfo {author}
  {\bibfnamefont {S.~Y.}\ \bibnamefont {Lee}}, \bibinfo {author} {\bibfnamefont
  {J.~M.}\ \bibnamefont {Kim}}, \bibinfo {author} {\bibfnamefont {K.~S.}\
  \bibnamefont {Kim}}, \bibinfo {author} {\bibfnamefont {J.-H.}\ \bibnamefont
  {Ahn}}, \bibinfo {author} {\bibfnamefont {P.}~\bibnamefont {Kim}}, \bibinfo
  {author} {\bibfnamefont {J.-Y.}\ \bibnamefont {Choi}}, \ and\ \bibinfo
  {author} {\bibfnamefont {B.~H.}\ \bibnamefont {Hong}},\ }\href@noop {}
  {\bibfield  {journal} {\bibinfo  {journal} {Nature}\ }\textbf {\bibinfo
  {volume} {457}},\ \bibinfo {pages} {706} (\bibinfo {year}
  {2008})}\BibitemShut {NoStop}%
\bibitem [{\citenamefont {Bonaccorso}\ \emph {et~al.}(2013)\citenamefont
  {Bonaccorso}, \citenamefont {Lombardo}, \citenamefont {Hasan}, \citenamefont
  {Sun}, \citenamefont {Colombo},\ and\ \citenamefont
  {Ferrari}}]{Bonaccorso:2013iy}%
  \BibitemOpen
  \bibfield  {author} {\bibinfo {author} {\bibfnamefont {F.}~\bibnamefont
  {Bonaccorso}}, \bibinfo {author} {\bibfnamefont {A.}~\bibnamefont
  {Lombardo}}, \bibinfo {author} {\bibfnamefont {T.}~\bibnamefont {Hasan}},
  \bibinfo {author} {\bibfnamefont {Z.}~\bibnamefont {Sun}}, \bibinfo {author}
  {\bibfnamefont {L.}~\bibnamefont {Colombo}}, \ and\ \bibinfo {author}
  {\bibfnamefont {A.~C.}\ \bibnamefont {Ferrari}},\ }\href@noop {} {\bibfield
  {journal} {\bibinfo  {journal} {Materials Today}\ }\textbf {\bibinfo {volume}
  {15}},\ \bibinfo {pages} {564} (\bibinfo {year} {2013})}\BibitemShut
  {NoStop}%
\bibitem [{\citenamefont {Zhao}\ \emph {et~al.}(2011)\citenamefont {Zhao},
  \citenamefont {Beekman}, \citenamefont {Sandilands}, \citenamefont
  {Bashucky}, \citenamefont {Kwok}, \citenamefont {Lee}, \citenamefont
  {LaForge}, \citenamefont {Cheong},\ and\ \citenamefont
  {Burch}}]{Zhao:2011bs}%
  \BibitemOpen
  \bibfield  {author} {\bibinfo {author} {\bibfnamefont {S.~Y.~F.}\
  \bibnamefont {Zhao}}, \bibinfo {author} {\bibfnamefont {C.}~\bibnamefont
  {Beekman}}, \bibinfo {author} {\bibfnamefont {L.~J.}\ \bibnamefont
  {Sandilands}}, \bibinfo {author} {\bibfnamefont {J.~E.~J.}\ \bibnamefont
  {Bashucky}}, \bibinfo {author} {\bibfnamefont {D.}~\bibnamefont {Kwok}},
  \bibinfo {author} {\bibfnamefont {N.}~\bibnamefont {Lee}}, \bibinfo {author}
  {\bibfnamefont {A.~D.}\ \bibnamefont {LaForge}}, \bibinfo {author}
  {\bibfnamefont {S.~W.}\ \bibnamefont {Cheong}}, \ and\ \bibinfo {author}
  {\bibfnamefont {K.~S.}\ \bibnamefont {Burch}},\ }\href@noop {} {\bibfield
  {journal} {\bibinfo  {journal} {Applied Physics Letters}\ }\textbf {\bibinfo
  {volume} {98}},\ \bibinfo {pages} {141911} (\bibinfo {year}
  {2011})}\BibitemShut {NoStop}%
\bibitem [{\citenamefont {Sandilands}\ \emph {et~al.}(2010)\citenamefont
  {Sandilands}, \citenamefont {Shen}, \citenamefont {Chugunov}, \citenamefont
  {Zhao}, \citenamefont {Ono}, \citenamefont {Ando},\ and\ \citenamefont
  {Burch}}]{Sandilands:2010tq}%
  \BibitemOpen
  \bibfield  {author} {\bibinfo {author} {\bibfnamefont {L.}~\bibnamefont
  {Sandilands}}, \bibinfo {author} {\bibfnamefont {J.}~\bibnamefont {Shen}},
  \bibinfo {author} {\bibfnamefont {G.}~\bibnamefont {Chugunov}}, \bibinfo
  {author} {\bibfnamefont {S.}~\bibnamefont {Zhao}}, \bibinfo {author}
  {\bibfnamefont {S.}~\bibnamefont {Ono}}, \bibinfo {author} {\bibfnamefont
  {Y.}~\bibnamefont {Ando}}, \ and\ \bibinfo {author} {\bibfnamefont
  {K.}~\bibnamefont {Burch}},\ }\href@noop {} {\bibfield  {journal} {\bibinfo
  {journal} {Physical Review B}\ }\textbf {\bibinfo {volume} {82}},\ \bibinfo
  {pages} {064503} (\bibinfo {year} {2010})}\BibitemShut {NoStop}%
\bibitem [{\citenamefont {LaForge}\ \emph {et~al.}(2010)\citenamefont
  {LaForge}, \citenamefont {Frenzel}, \citenamefont {Pursley}, \citenamefont
  {Lin}, \citenamefont {Liu}, \citenamefont {Shi},\ and\ \citenamefont
  {Basov}}]{LaForge:2010dx}%
  \BibitemOpen
  \bibfield  {author} {\bibinfo {author} {\bibfnamefont {A.~D.}\ \bibnamefont
  {LaForge}}, \bibinfo {author} {\bibfnamefont {A.}~\bibnamefont {Frenzel}},
  \bibinfo {author} {\bibfnamefont {B.~C.}\ \bibnamefont {Pursley}}, \bibinfo
  {author} {\bibfnamefont {T.}~\bibnamefont {Lin}}, \bibinfo {author}
  {\bibfnamefont {X.}~\bibnamefont {Liu}}, \bibinfo {author} {\bibfnamefont
  {J.}~\bibnamefont {Shi}}, \ and\ \bibinfo {author} {\bibfnamefont {D.~N.}\
  \bibnamefont {Basov}},\ }\href@noop {} {\bibfield  {journal} {\bibinfo
  {journal} {Physical Review B}\ }\textbf {\bibinfo {volume} {81}},\ \bibinfo
  {pages} {125120} (\bibinfo {year} {2010})}\BibitemShut {NoStop}%
\bibitem [{\citenamefont {Tran}\ \emph {et~al.}(2014)\citenamefont {Tran},
  \citenamefont {Levallois}, \citenamefont {Lerch}, \citenamefont {Teyssier},
  \citenamefont {Kuzmenko}, \citenamefont {Aut{\`e}s}, \citenamefont {Yazyev},
  \citenamefont {Ubaldini}, \citenamefont {Giannini}, \citenamefont {van~der
  Marel},\ and\ \citenamefont {Akrap}}]{Tran:2014if}%
  \BibitemOpen
  \bibfield  {author} {\bibinfo {author} {\bibfnamefont {M.~K.}\ \bibnamefont
  {Tran}}, \bibinfo {author} {\bibfnamefont {J.}~\bibnamefont {Levallois}},
  \bibinfo {author} {\bibfnamefont {P.}~\bibnamefont {Lerch}}, \bibinfo
  {author} {\bibfnamefont {J.}~\bibnamefont {Teyssier}}, \bibinfo {author}
  {\bibfnamefont {A.~B.}\ \bibnamefont {Kuzmenko}}, \bibinfo {author}
  {\bibfnamefont {G.}~\bibnamefont {Aut{\`e}s}}, \bibinfo {author}
  {\bibfnamefont {O.~V.}\ \bibnamefont {Yazyev}}, \bibinfo {author}
  {\bibfnamefont {A.}~\bibnamefont {Ubaldini}}, \bibinfo {author}
  {\bibfnamefont {E.}~\bibnamefont {Giannini}}, \bibinfo {author}
  {\bibfnamefont {D.}~\bibnamefont {van~der Marel}}, \ and\ \bibinfo {author}
  {\bibfnamefont {A.}~\bibnamefont {Akrap}},\ }\href@noop {} {\bibfield
  {journal} {\bibinfo  {journal} {Physical Review Letters}\ }\textbf {\bibinfo
  {volume} {112}},\ \bibinfo {pages} {047402} (\bibinfo {year}
  {2014})}\BibitemShut {NoStop}%
\bibitem [{\citenamefont {Xiong}\ \emph
  {et~al.}(2012{\natexlab{a}})\citenamefont {Xiong}, \citenamefont {Petersen},
  \citenamefont {Qu}, \citenamefont {Hor}, \citenamefont {Cava},\ and\
  \citenamefont {Ong}}]{Xiong:2011wn}%
  \BibitemOpen
  \bibfield  {author} {\bibinfo {author} {\bibfnamefont {J.}~\bibnamefont
  {Xiong}}, \bibinfo {author} {\bibfnamefont {A.~C.}\ \bibnamefont {Petersen}},
  \bibinfo {author} {\bibfnamefont {D.}~\bibnamefont {Qu}}, \bibinfo {author}
  {\bibfnamefont {Y.~S.}\ \bibnamefont {Hor}}, \bibinfo {author} {\bibfnamefont
  {R.~J.}\ \bibnamefont {Cava}}, \ and\ \bibinfo {author} {\bibfnamefont
  {N.~P.}\ \bibnamefont {Ong}},\ }\href@noop {} {\bibfield  {journal} {\bibinfo
   {journal} {Physica E: Low-dimensional Systems and Nanostructures}\ }\textbf
  {\bibinfo {volume} {44}},\ \bibinfo {pages} {917} (\bibinfo {year}
  {2012}{\natexlab{a}})}\BibitemShut {NoStop}%
\bibitem [{\citenamefont {Forro}\ \emph {et~al.}(1990)\citenamefont {Forro},
  \citenamefont {Carr}, \citenamefont {Williams}, \citenamefont {Mandrus},\
  and\ \citenamefont {Mihaly}}]{Forro:1990vt}%
  \BibitemOpen
  \bibfield  {author} {\bibinfo {author} {\bibfnamefont {L.}~\bibnamefont
  {Forro}}, \bibinfo {author} {\bibfnamefont {G.~L.}\ \bibnamefont {Carr}},
  \bibinfo {author} {\bibfnamefont {G.~P.}\ \bibnamefont {Williams}}, \bibinfo
  {author} {\bibfnamefont {D.}~\bibnamefont {Mandrus}}, \ and\ \bibinfo
  {author} {\bibfnamefont {L.}~\bibnamefont {Mihaly}},\ }\href@noop {}
  {\bibfield  {journal} {\bibinfo  {journal} {Physical Review Letters}\
  }\textbf {\bibinfo {volume} {65}},\ \bibinfo {pages} {1941} (\bibinfo {year}
  {1990})}\BibitemShut {NoStop}%
\bibitem [{\citenamefont {Frindt}(1965)}]{Frindt:1965to}%
  \BibitemOpen
  \bibfield  {author} {\bibinfo {author} {\bibfnamefont {R.~F.}\ \bibnamefont
  {Frindt}},\ }\href@noop {} {\bibfield  {journal} {\bibinfo  {journal}
  {Physical Review}\ }\textbf {\bibinfo {volume} {140}},\ \bibinfo {pages}
  {A536} (\bibinfo {year} {1965})}\BibitemShut {NoStop}%
\bibitem [{\citenamefont {Frindt}(1972)}]{1972PhRvL..28..299F}%
  \BibitemOpen
  \bibfield  {author} {\bibinfo {author} {\bibfnamefont {R.~F.}\ \bibnamefont
  {Frindt}},\ }\href@noop {} {\bibfield  {journal} {\bibinfo  {journal}
  {Physical Review Letters}\ }\textbf {\bibinfo {volume} {28}},\ \bibinfo
  {pages} {299} (\bibinfo {year} {1972})}\BibitemShut {NoStop}%
\bibitem [{\citenamefont {Hecht}(2002)}]{hecht}%
  \BibitemOpen
  \bibfield  {author} {\bibinfo {author} {\bibfnamefont {E.}~\bibnamefont
  {Hecht}},\ }\href@noop {} {\emph {\bibinfo {title} {Optics}}},\ \bibinfo
  {edition} {4th}\ ed.\ (\bibinfo  {publisher} {Addison Wesley},\ \bibinfo
  {year} {2002})\BibitemShut {NoStop}%
\bibitem [{\citenamefont {Reijnders}\ \emph {et~al.}(2014)\citenamefont
  {Reijnders}, \citenamefont {Tian}, \citenamefont {Sandilands}, \citenamefont
  {Pohl}, \citenamefont {Kivlichan}, \citenamefont {Frank~Zhao}, \citenamefont
  {Jia}, \citenamefont {Charles}, \citenamefont {Cava}, \citenamefont
  {Alidoust}, \citenamefont {Xu}, \citenamefont {Neupane}, \citenamefont
  {Hasan}, \citenamefont {Wang}, \citenamefont {Cheong},\ and\ \citenamefont
  {Burch}}]{Reijnders:2014hq}%
  \BibitemOpen
  \bibfield  {author} {\bibinfo {author} {\bibfnamefont {A.}~\bibnamefont
  {Reijnders}}, \bibinfo {author} {\bibfnamefont {Y.}~\bibnamefont {Tian}},
  \bibinfo {author} {\bibfnamefont {L.~J.}\ \bibnamefont {Sandilands}},
  \bibinfo {author} {\bibfnamefont {G.}~\bibnamefont {Pohl}}, \bibinfo {author}
  {\bibfnamefont {I.~D.}\ \bibnamefont {Kivlichan}}, \bibinfo {author}
  {\bibfnamefont {S.~Y.}\ \bibnamefont {Frank~Zhao}}, \bibinfo {author}
  {\bibfnamefont {S.}~\bibnamefont {Jia}}, \bibinfo {author} {\bibfnamefont
  {M.~E.}\ \bibnamefont {Charles}}, \bibinfo {author} {\bibfnamefont {R.~J.}\
  \bibnamefont {Cava}}, \bibinfo {author} {\bibfnamefont {N.}~\bibnamefont
  {Alidoust}}, \bibinfo {author} {\bibfnamefont {S.}~\bibnamefont {Xu}},
  \bibinfo {author} {\bibfnamefont {M.}~\bibnamefont {Neupane}}, \bibinfo
  {author} {\bibfnamefont {M.~Z.}\ \bibnamefont {Hasan}}, \bibinfo {author}
  {\bibfnamefont {X.}~\bibnamefont {Wang}}, \bibinfo {author} {\bibfnamefont
  {S.~W.}\ \bibnamefont {Cheong}}, \ and\ \bibinfo {author} {\bibfnamefont
  {K.~S.}\ \bibnamefont {Burch}},\ }\href@noop {} {\bibfield  {journal}
  {\bibinfo  {journal} {Physical Review B}\ }\textbf {\bibinfo {volume} {89}},\
  \bibinfo {pages} {075138} (\bibinfo {year} {2014})}\BibitemShut {NoStop}%
\bibitem [{\citenamefont {Palik}(1985)}]{Palik}%
  \BibitemOpen
  \bibfield  {author} {\bibinfo {author} {\bibfnamefont {E.~D.}\ \bibnamefont
  {Palik}},\ }\href@noop {} {\emph {\bibinfo {title} {Handbook of Optical
  Constants of Solids,}}}\ (\bibinfo  {publisher} {Academic Press, Boston},\
  \bibinfo {year} {1985})\BibitemShut {NoStop}%
\bibitem [{\citenamefont {{Kuzmenko, A. B.}}(2005)}]{KuzmenkoAB:2005jh}%
  \BibitemOpen
  \bibfield  {author} {\bibinfo {author} {\bibnamefont {{Kuzmenko, A. B.}}},\
  }\href@noop {} {\bibfield  {journal} {\bibinfo  {journal} {Rev. Sci.
  Instrum.}\ }\textbf {\bibinfo {volume} {76}},\ \bibinfo {pages} {083108}
  (\bibinfo {year} {2005})}\BibitemShut {NoStop}%
\bibitem [{\citenamefont {Blom}\ \emph {et~al.}(1997)\citenamefont {Blom},
  \citenamefont {van Dijk}, \citenamefont {Hoekstra}, \citenamefont
  {Driessen},\ and\ \citenamefont {Popma}}]{Blom:1997fj}%
  \BibitemOpen
  \bibfield  {author} {\bibinfo {author} {\bibfnamefont {F.~C.}\ \bibnamefont
  {Blom}}, \bibinfo {author} {\bibfnamefont {D.~R.}\ \bibnamefont {van Dijk}},
  \bibinfo {author} {\bibfnamefont {H.~J. W.~M.}\ \bibnamefont {Hoekstra}},
  \bibinfo {author} {\bibfnamefont {A.}~\bibnamefont {Driessen}}, \ and\
  \bibinfo {author} {\bibfnamefont {T.~J.~A.}\ \bibnamefont {Popma}},\
  }\href@noop {} {\bibfield  {journal} {\bibinfo  {journal} {Applied Physics
  Letters}\ }\textbf {\bibinfo {volume} {71}},\ \bibinfo {pages} {747}
  (\bibinfo {year} {1997})}\BibitemShut {NoStop}%
\bibitem [{\citenamefont {Cooley}, \citenamefont {Lewis},\ and\ \citenamefont
  {Welch}(1969)}]{1969ITEdu..12...27C}%
  \BibitemOpen
  \bibfield  {author} {\bibinfo {author} {\bibfnamefont {J.~W.}\ \bibnamefont
  {Cooley}}, \bibinfo {author} {\bibfnamefont {P.~A.~W.}\ \bibnamefont
  {Lewis}}, \ and\ \bibinfo {author} {\bibfnamefont {P.~D.}\ \bibnamefont
  {Welch}},\ }\href@noop {} {\bibfield  {journal} {\bibinfo  {journal} {IEEE
  Transactions on Education}\ }\textbf {\bibinfo {volume} {12}},\ \bibinfo
  {pages} {27} (\bibinfo {year} {1969})}\BibitemShut {NoStop}%
\bibitem [{Pol(2010)}]{Polypropylene}%
  \BibitemOpen
  \href@noop {} {\emph {\bibinfo {title} {Refractive index for Polypropylene
  from manufacturers specifications}}} (\bibinfo {year} {2010})\BibitemShut
  {NoStop}%
\bibitem [{\citenamefont {Birch}(1992)}]{Birch:1992uc}%
  \BibitemOpen
  \bibfield  {author} {\bibinfo {author} {\bibfnamefont {J.~R.}\ \bibnamefont
  {Birch}},\ }\href@noop {} {\bibfield  {journal} {\bibinfo  {journal}
  {Infrared physics}\ }\textbf {\bibinfo {volume} {33}},\ \bibinfo {pages} {33}
  (\bibinfo {year} {1992})}\BibitemShut {NoStop}%
\bibitem [{\citenamefont {Davis}, \citenamefont {Abrams},\ and\ \citenamefont
  {Brault}(2001)}]{davis2001fourier}%
  \BibitemOpen
  \bibfield  {author} {\bibinfo {author} {\bibfnamefont {S.}~\bibnamefont
  {Davis}}, \bibinfo {author} {\bibfnamefont {M.}~\bibnamefont {Abrams}}, \
  and\ \bibinfo {author} {\bibfnamefont {J.}~\bibnamefont {Brault}},\ }\href
  {http://books.google.ca/books?id=gMXrvGg5hTQC} {\emph {\bibinfo {title}
  {Fourier Transform Spectrometry}}}\ (\bibinfo  {publisher} {Elsevier
  Science},\ \bibinfo {year} {2001})\BibitemShut {NoStop}%
\bibitem [{\citenamefont {Segura}\ \emph {et~al.}(2012)\citenamefont {Segura},
  \citenamefont {Panchal}, \citenamefont {S{\'a}nchez-Royo}, \citenamefont
  {Mar{\'\i}n-Borr{\'a}s}, \citenamefont {Munoz-Sanjos{\'e}}, \citenamefont
  {Rodr{\'\i}guez-Hern{\'a}ndez}, \citenamefont {Munoz}, \citenamefont
  {P{\'e}rez-Gonz{\'a}lez}, \citenamefont {Manj{\'o}n},\ and\ \citenamefont
  {Gonz{\'a}lez}}]{Segura:2012wb}%
  \BibitemOpen
  \bibfield  {author} {\bibinfo {author} {\bibfnamefont {A.}~\bibnamefont
  {Segura}}, \bibinfo {author} {\bibfnamefont {V.}~\bibnamefont {Panchal}},
  \bibinfo {author} {\bibfnamefont {J.~F.}\ \bibnamefont {S{\'a}nchez-Royo}},
  \bibinfo {author} {\bibfnamefont {V.}~\bibnamefont {Mar{\'\i}n-Borr{\'a}s}},
  \bibinfo {author} {\bibfnamefont {V.}~\bibnamefont {Munoz-Sanjos{\'e}}},
  \bibinfo {author} {\bibfnamefont {P.}~\bibnamefont
  {Rodr{\'\i}guez-Hern{\'a}ndez}}, \bibinfo {author} {\bibfnamefont
  {A.}~\bibnamefont {Munoz}}, \bibinfo {author} {\bibfnamefont
  {E.}~\bibnamefont {P{\'e}rez-Gonz{\'a}lez}}, \bibinfo {author} {\bibfnamefont
  {F.~J.}\ \bibnamefont {Manj{\'o}n}}, \ and\ \bibinfo {author} {\bibfnamefont
  {J.}~\bibnamefont {Gonz{\'a}lez}},\ }\href@noop {} {\bibfield  {journal}
  {\bibinfo  {journal} {Physical Review B}\ }\textbf {\bibinfo {volume} {85}},\
  \bibinfo {pages} {195139} (\bibinfo {year} {2012})}\BibitemShut {NoStop}%
\bibitem [{\citenamefont {Akrap}\ \emph {et~al.}(2012)\citenamefont {Akrap},
  \citenamefont {Tran}, \citenamefont {Ubaldini}, \citenamefont {Teyssier},
  \citenamefont {Giannini}, \citenamefont {van~der Marel}, \citenamefont
  {Lerch},\ and\ \citenamefont {Homes}}]{2012arXiv1209.3593A}%
  \BibitemOpen
  \bibfield  {author} {\bibinfo {author} {\bibfnamefont {A.}~\bibnamefont
  {Akrap}}, \bibinfo {author} {\bibfnamefont {M.}~\bibnamefont {Tran}},
  \bibinfo {author} {\bibfnamefont {A.}~\bibnamefont {Ubaldini}}, \bibinfo
  {author} {\bibfnamefont {J.}~\bibnamefont {Teyssier}}, \bibinfo {author}
  {\bibfnamefont {E.}~\bibnamefont {Giannini}}, \bibinfo {author}
  {\bibfnamefont {D.}~\bibnamefont {van~der Marel}}, \bibinfo {author}
  {\bibfnamefont {P.}~\bibnamefont {Lerch}}, \ and\ \bibinfo {author}
  {\bibfnamefont {C.~C.}\ \bibnamefont {Homes}},\ }\href@noop {} {\bibfield
  {journal} {\bibinfo  {journal} {Physical Review B}\ }\textbf {\bibinfo
  {volume} {86}},\ \bibinfo {pages} {235207} (\bibinfo {year}
  {2012})}\BibitemShut {NoStop}%
\bibitem [{\citenamefont {Xiong}\ \emph
  {et~al.}(2012{\natexlab{b}})\citenamefont {Xiong}, \citenamefont {Luo},
  \citenamefont {Khoo}, \citenamefont {Jia}, \citenamefont {Cava},\ and\
  \citenamefont {Ong}}]{Xiong:2011tm}%
  \BibitemOpen
  \bibfield  {author} {\bibinfo {author} {\bibfnamefont {J.}~\bibnamefont
  {Xiong}}, \bibinfo {author} {\bibfnamefont {Y.}~\bibnamefont {Luo}}, \bibinfo
  {author} {\bibfnamefont {Y.}~\bibnamefont {Khoo}}, \bibinfo {author}
  {\bibfnamefont {S.}~\bibnamefont {Jia}}, \bibinfo {author} {\bibfnamefont
  {R.~J.}\ \bibnamefont {Cava}}, \ and\ \bibinfo {author} {\bibfnamefont
  {N.~P.}\ \bibnamefont {Ong}},\ }\href@noop {} {\bibfield  {journal} {\bibinfo
   {journal} {Physical Review B}\ }\textbf {\bibinfo {volume} {86}},\ \bibinfo
  {pages} {45314} (\bibinfo {year} {2012}{\natexlab{b}})}\BibitemShut {NoStop}%
\bibitem [{\citenamefont {Ji}\ \emph {et~al.}(2013)\citenamefont {Ji},
  \citenamefont {Stokes}, \citenamefont {Alegria}, \citenamefont {Blomberg},
  \citenamefont {Tanata}, \citenamefont {Reijnders}, \citenamefont {Schoop},
  \citenamefont {Tian}, \citenamefont {Prozorov}, \citenamefont {Burch},
  \citenamefont {Ong}, \citenamefont {Petta},\ and\ \citenamefont
  {Cava}}]{Ji:2013dg}%
  \BibitemOpen
  \bibfield  {author} {\bibinfo {author} {\bibfnamefont {H.}~\bibnamefont
  {Ji}}, \bibinfo {author} {\bibfnamefont {R.~A.}\ \bibnamefont {Stokes}},
  \bibinfo {author} {\bibfnamefont {L.~D.}\ \bibnamefont {Alegria}}, \bibinfo
  {author} {\bibfnamefont {E.~C.}\ \bibnamefont {Blomberg}}, \bibinfo {author}
  {\bibfnamefont {M.~A.}\ \bibnamefont {Tanata}}, \bibinfo {author}
  {\bibfnamefont {A.}~\bibnamefont {Reijnders}}, \bibinfo {author}
  {\bibfnamefont {L.~M.}\ \bibnamefont {Schoop}}, \bibinfo {author}
  {\bibfnamefont {L.}~\bibnamefont {Tian}}, \bibinfo {author} {\bibfnamefont
  {R.}~\bibnamefont {Prozorov}}, \bibinfo {author} {\bibfnamefont {K.~S.}\
  \bibnamefont {Burch}}, \bibinfo {author} {\bibfnamefont {N.~P.}\ \bibnamefont
  {Ong}}, \bibinfo {author} {\bibfnamefont {J.~R.}\ \bibnamefont {Petta}}, \
  and\ \bibinfo {author} {\bibfnamefont {R.~J.}\ \bibnamefont {Cava}},\ }\href
  {http://scitation.aip.org/getpdf/servlet/GetPDFServlet?filetype=pdf&id=JAPIAU000114000011114907000001&idtype=cvips&doi=10.1063/1.4822092&prog=normal}
  {\bibfield  {journal} {\bibinfo  {journal} {Journa of Applied Physics}\
  }\textbf {\bibinfo {volume} {114}},\ \bibinfo {pages} {114907} (\bibinfo
  {year} {2013})}\BibitemShut {NoStop}%
\bibitem [{\citenamefont {Plumb}\ \emph {et~al.}(2014)\citenamefont {Plumb},
  \citenamefont {Clancy}, \citenamefont {Sandilands}, \citenamefont {Shankar},
  \citenamefont {Hu}, \citenamefont {Burch}, \citenamefont {Kee},\ and\
  \citenamefont {Kim}}]{Plumb:2014tv}%
  \BibitemOpen
  \bibfield  {author} {\bibinfo {author} {\bibfnamefont {K.~W.}\ \bibnamefont
  {Plumb}}, \bibinfo {author} {\bibfnamefont {J.~P.}\ \bibnamefont {Clancy}},
  \bibinfo {author} {\bibfnamefont {L.}~\bibnamefont {Sandilands}}, \bibinfo
  {author} {\bibfnamefont {V.~V.}\ \bibnamefont {Shankar}}, \bibinfo {author}
  {\bibfnamefont {Y.~F.}\ \bibnamefont {Hu}}, \bibinfo {author} {\bibfnamefont
  {K.~S.}\ \bibnamefont {Burch}}, \bibinfo {author} {\bibfnamefont {H.-Y.}\
  \bibnamefont {Kee}}, \ and\ \bibinfo {author} {\bibfnamefont {Y.-J.}\
  \bibnamefont {Kim}},\ }\href@noop {} {\bibfield  {journal} {\bibinfo
  {journal} {arXiv}\ ,\ \bibinfo {pages} {1403.0883v2}} (\bibinfo {year}
  {2014})}\BibitemShut {NoStop}%
\bibitem [{\citenamefont {Reijnders}\ and\ \citenamefont
  {Burch}(2014)}]{Reijnders3}%
  \BibitemOpen
  \bibfield  {author} {\bibinfo {author} {\bibfnamefont {A.}~\bibnamefont
  {Reijnders}}\ and\ \bibinfo {author} {\bibfnamefont {K.~S.}\ \bibnamefont
  {Burch}},\ }\href@noop {} {\bibfield  {journal} {\bibinfo  {journal} {(To be
  published)}\ } (\bibinfo {year} {2014})}\BibitemShut {NoStop}%
\bibitem [{\citenamefont {Sandilands}\ \emph {et~al.}(2014)\citenamefont
  {Sandilands}, \citenamefont {Reijnders}, \citenamefont {Plumb}, \citenamefont
  {Kim},\ and\ \citenamefont {Burch}}]{Sandilands3}%
  \BibitemOpen
  \bibfield  {author} {\bibinfo {author} {\bibfnamefont {L.~J.}\ \bibnamefont
  {Sandilands}}, \bibinfo {author} {\bibfnamefont {A.~A.}\ \bibnamefont
  {Reijnders}}, \bibinfo {author} {\bibfnamefont {K.~W.}\ \bibnamefont
  {Plumb}}, \bibinfo {author} {\bibfnamefont {Y.-J.}\ \bibnamefont {Kim}}, \
  and\ \bibinfo {author} {\bibfnamefont {K.~S.}\ \bibnamefont {Burch}},\
  }\href@noop {} {\bibfield  {journal} {\bibinfo  {journal} {(To be
  published)}\ } (\bibinfo {year} {2014})}\BibitemShut {NoStop}%
\bibitem [{\citenamefont {Kohler}\ and\ \citenamefont
  {Becker}(1974)}]{KOHLER:1974tf}%
  \BibitemOpen
  \bibfield  {author} {\bibinfo {author} {\bibfnamefont {H.}~\bibnamefont
  {Kohler}}\ and\ \bibinfo {author} {\bibfnamefont {C.}~\bibnamefont
  {Becker}},\ }\href@noop {} {\bibfield  {journal} {\bibinfo  {journal}
  {Physica Status Solidi B}\ }\textbf {\bibinfo {volume} {61}},\ \bibinfo
  {pages} {533} (\bibinfo {year} {1974})}\BibitemShut {NoStop}%
\bibitem [{\citenamefont {Jenkins}\ \emph {et~al.}(2010)\citenamefont
  {Jenkins}, \citenamefont {Sushkov}, \citenamefont {Schmadel}, \citenamefont
  {Butch}, \citenamefont {Syers}, \citenamefont {Paglione},\ and\ \citenamefont
  {Drew}}]{Jenkins:2010kg}%
  \BibitemOpen
  \bibfield  {author} {\bibinfo {author} {\bibfnamefont {G.}~\bibnamefont
  {Jenkins}}, \bibinfo {author} {\bibfnamefont {A.}~\bibnamefont {Sushkov}},
  \bibinfo {author} {\bibfnamefont {D.}~\bibnamefont {Schmadel}}, \bibinfo
  {author} {\bibfnamefont {N.}~\bibnamefont {Butch}}, \bibinfo {author}
  {\bibfnamefont {P.}~\bibnamefont {Syers}}, \bibinfo {author} {\bibfnamefont
  {J.}~\bibnamefont {Paglione}}, \ and\ \bibinfo {author} {\bibfnamefont
  {H.}~\bibnamefont {Drew}},\ }\href@noop {} {\bibfield  {journal} {\bibinfo
  {journal} {Physical Review B}\ }\textbf {\bibinfo {volume} {82}},\ \bibinfo
  {pages} {125120} (\bibinfo {year} {2010})}\BibitemShut {NoStop}%
\bibitem [{\citenamefont {Cao}\ \emph {et~al.}(2012)\citenamefont {Cao},
  \citenamefont {Tian}, \citenamefont {Miotkowski}, \citenamefont {Shen},
  \citenamefont {Hu}, \citenamefont {Qiao},\ and\ \citenamefont
  {Chen}}]{Cao:2012kc}%
  \BibitemOpen
  \bibfield  {author} {\bibinfo {author} {\bibfnamefont {H.}~\bibnamefont
  {Cao}}, \bibinfo {author} {\bibfnamefont {J.}~\bibnamefont {Tian}}, \bibinfo
  {author} {\bibfnamefont {I.}~\bibnamefont {Miotkowski}}, \bibinfo {author}
  {\bibfnamefont {T.}~\bibnamefont {Shen}}, \bibinfo {author} {\bibfnamefont
  {J.}~\bibnamefont {Hu}}, \bibinfo {author} {\bibfnamefont {S.}~\bibnamefont
  {Qiao}}, \ and\ \bibinfo {author} {\bibfnamefont {Y.~P.}\ \bibnamefont
  {Chen}},\ }\href@noop {} {\bibfield  {journal} {\bibinfo  {journal} {Physical
  Review Letters}\ }\textbf {\bibinfo {volume} {108}},\ \bibinfo {pages}
  {216803} (\bibinfo {year} {2012})}\BibitemShut {NoStop}%
\end{thebibliography}
%

\end{document}